\documentclass[a4paper,twocolumn,english]{revtex4}
\usepackage[T1]{fontenc}
\usepackage[latin9]{inputenc}
\usepackage{amsthm}
\usepackage{amsmath}
\usepackage{graphicx}
\usepackage{esint}

\makeatletter

\pdfpageheight\paperheight
\pdfpagewidth\paperwidth

\DeclareRobustCommand{\cyrtext}{%
  \fontencoding{T2A}\selectfont\def\encodingdefault{T2A}}
\DeclareRobustCommand{\textcyr}[1]{\leavevmode{\cyrtext #1}}
\AtBeginDocument{\DeclareFontEncoding{T2A}{}{}}

\@ifundefined{textcolor}{}
{%
 \definecolor{BLACK}{gray}{0}
 \definecolor{WHITE}{gray}{1}
 \definecolor{RED}{rgb}{1,0,0}
 \definecolor{GREEN}{rgb}{0,1,0}
 \definecolor{BLUE}{rgb}{0,0,1}
 \definecolor{CYAN}{cmyk}{1,0,0,0}
 \definecolor{MAGENTA}{cmyk}{0,1,0,0}
 \definecolor{YELLOW}{cmyk}{0,0,1,0}
}

\makeatother

\usepackage{babel}
\begin{document}

\title{Wetting regimes and interactions of parallel plane surfaces in \textcyr{\char224}
polar liquid}

\author{P. O. Fedichev$^{1,2}$ L. I. Menshikov$^{1,3}$ }

\affiliation{$^{1)}$Quantum Pharmaceuticals Ltd, Ul. Kosmonavta Volkova 6A-606,
125171, Moscow, Russian Federation}

\affiliation{$^{2)}$Moscow Institute of Physics and Technology, 141700, Institutskii
per. 9, Dolgoprudny, Moscow Region, Russian Federation}

\affiliation{$^{3)}$RRC Kurchatov Institute, Kurchatov Square 1, 123182, Moscow,
Russian Federation}

\email{peter.fedichev@q-pharm.com}

\homepage{http://q-pharm.com}

\begin{abstract}
We apply a phenomenological theory of polar liquids to calculate the
interaction energy between two plane surfaces at $nm$-distances.
We show that depending on the properties of the surface-liquid interfaces,
the interacting surfaces induce polarization of the liquid in different
ways. We find, in full agreement with available experiments, that
if the interfaces are mostly hydrophobic, then the interaction is
attractive and relatively long-ranged (interaction decay length $\lambda\sim1.2\, nm$).
The water molecules are net polarized parallel to the surfaces in
this case. If the surfaces are mostly hydrophilic, then the molecules
are polarized against the surfaces, and the interaction becomes repulsive,
but at a short-range ($\lambda\sim0.2\, nm$). Finally, we predict
there exists an intermediate regime, where the surfaces fail to order
the water molecules, the interaction becomes much weaker, attractive
and, at relatively small distances, decays with the inverse square
of the distance between the surfaces. 
\end{abstract}
\maketitle
Interaction forces between hydrated, $nm$-size objects at short distances
play an important role in various biological and nano-fabrication
processes. For example, the disjointing pressure between two biological
membranes in pure water at distances $0.5\, nm\alt d\alt2.5\, nm$
corresponds to a short-range repulsive force 
\begin{equation}
P=-S^{-1}\partial G/\partial d=P_{0}\exp(-d/\lambda),\label{eq: The disjoining pressure}
\end{equation}
where $S$ is the cross-sectional area, $G(d)$ is the interaction
energy of the system, $5\cdot10^{9}\, dyn/cm^{2}\alt P_{0}\alt5\cdot10^{10}\, dyn/cm^{2}$
, and $\lambda\sim0.2\, nm$ \cite{leneveu1976measurement}. On the
other hand, two hydrophobic plane surfaces exhibit attraction, $P=-P_{0}^{\prime}\exp\left(-d/\lambda^{\prime}\right)$,
characterized by a similar prefactor, $P_{0}^{\prime}\sim P_{0}$,
but different exponent: $\lambda^{\prime}\sim1.2\, nm$ \cite{pashley1985attractive,claesson1988very}.
Therefore, experiments show that hydration forces are characterized
by at least two different length scales and depend on surface material
properties.

The nature of these forces are explained by a few theoretical approaches.
The Landau-type model with the order parameter corresponding to the
ordering of the water molecules was presented in \cite{marcelja1976repulsion}
to describe the repulsion of hydrophilic surfaces. In a related approach,
this order parameter was instead associated with the hydrogen-bond
network deformations \cite{kjellander1985inhomogeneous}. In \cite{belaya1986hydration,kornyshev1989fluctuation,leikin1990theory},
the hydration forces between phospholipid membranes were associated
with non-local polarization of the liquid. All of the models are purely
phenomenological and provide an explanation for the repulsion force
(\ref{eq: The disjoining pressure}), although the specific values
for the parameters $P_{1}$ and $\lambda$ cannot be established from
the theory. 

In this letter, we use a previously developed phenomenological model
\cite{fedichev2006long,men2011possible,men2009nature} of a polar
liquid to describe the interaction between plane surfaces arbitrary
interface properties. We characterize the liquid by the microscopic
average of the molecular dipole moment orientations vector, $\mathbf{s}(\mathbf{r})$,
over a microscopic volume element centered at position $\mathbf{r}$
that contains a macroscopically large number of molecules. In the
following, we represent the free energy of the liquid as $G[\mathbf{s\left(\mathbf{r}\right)}]=G_{B}+G_{S},$
where $G_{S}$ is the energy of the liquid-surface interface (see
below), and 
\[
G_{B}[\mathbf{s\left(\mathbf{r}\right)}]=P_{0}^{2}\int dV\frac{C}{2}\sum_{\alpha,\beta=x,y,z}\frac{\partial s_{\alpha}}{\partial x_{\beta}}\frac{\partial s_{\alpha}}{\partial x_{\beta}}+
\]
\begin{equation}
+\int dV\left[P_{0}^{2}V(\mathbf{s}^{2})+{\displaystyle \frac{1}{8\pi}\mathbf{E}_{P}^{2}}-\mathbf{P\left(\mathbf{r}\right)E_{\mathbf{e}}\left(\mathbf{r}\right)}\right],\label{eq:GB}
\end{equation}
is the energy of the liquid bulk. Here, $P_{0}=n_{0}d_{0}$, $n_{0}$
is the density of the liquid, $d_{0}$ is the molecular dipole momentum
and the Oseen energy term is characterized by $C\approx0.5\, nm^{2}$,
the phenomenological constant responsible for the short range hydrogen
bonds stiffness. The polarization density of the liquid, $\mathbf{P}=P_{0}\mathbf{s}$
is related to the density of the polarization charges, $\rho_{P}=-{\rm div}\mathbf{P}$.
The polarization electric field is the solution of the Poisson equation
$\mathbf{{\rm div}E}_{P}=-4\pi\rho_{P}$, and $\mathbf{E}_{e}$ is
the external electric field in the absence of the liquid.

When the liquid polarization is small, $s\ll1$, ``the equation of
state'' function takes the usual Ginzburg-Landau form $V(\mathbf{s}^{2})\approx As^{2}/2+Bs^{4}$,
where $A$ and $B$ are the phenomenological liquid-dependent constants.
The former is determined by the long-range dipole-dipole interactions
in the liquid and is related to dielectric constant, $A=4\pi/\left(\varepsilon-1\right)$.
Water is characterized by a large value of $\varepsilon\approx80\gg1$.
Therefore, $A\approx0.16$, depends on the temperature and includes
the entropy contribution arising due to the averaging over the molecular
orientation. The smallness of $A$ is related to proximity of ferro-electric
phase, $A\approx\left(T-T_{C}\right)/T_{C}$, predicted by the model
and recently observed at temperatures $T=T_{C}\approx228K$ \cite{men2011possible,fedichev2011experimental}.
On the contrary, the parameter $B$ depends on the short-range physics
only, and is practically temperature independent, $B\sim1$. The thermal
state of bulk water in the model is characterized by the two scales:
$R_{D}=\sqrt{C/\left(4\pi+A\right)},\,0.15\, nm\leq R_{D}\leq0.25\, nm,$
is the size of the strongly correlated molecular cluster, and $L_{T}=\sqrt{C/A}\approx R_{D}\sqrt{\varepsilon},\,1.1\, nm\leq L_{T}\leq1.5\, nm$
($L_{T}\gg R_{D}$), is the size of the largest correlated domain
within the liquid. 

The interaction of the liquid and an immersed body surface is described
by \cite{men2009nature}:

\begin{equation}
G_{S}=-\frac{1}{2}\sqrt{C}P_{0}^{2}\int_{\Gamma}df\left(\alpha_{0}s_{\parallel}^{2}+\beta_{0}s_{\perp}^{2}\right),\label{eq:GS}
\end{equation}
where $df$ is the area element of the interface surface $\Gamma$,
the projections $s_{\perp}=\boldsymbol{n}\boldsymbol{s}$ and $\boldsymbol{s}_{\parallel}=\boldsymbol{s}-s_{\perp}\boldsymbol{n}$
are the normal and tangent components of the vector $\boldsymbol{s}$
and $\boldsymbol{n}$ is the unit vector normal to the interface surface,
directed into water bulk. The dimensionless phenomenological constants,
$\alpha_{0}$ and $\beta_{0}$, characterize the orientation dependent
interaction of the water molecules with the interface. These parameters
are liquid and surface material specific, and should be found from
either experimental data or molecular dynamics simulations. Once all
these parameters are known, the minimization of total functional $G$
with respect to the independent variable $\boldsymbol{s}\left(\boldsymbol{r}\right)$
leads to the Euler equation in the bulk and provides the proper boundary
condition for the vector field $\boldsymbol{s}\left(\boldsymbol{r}\right)$.

\begin{figure}
\includegraphics[width=0.9\columnwidth]{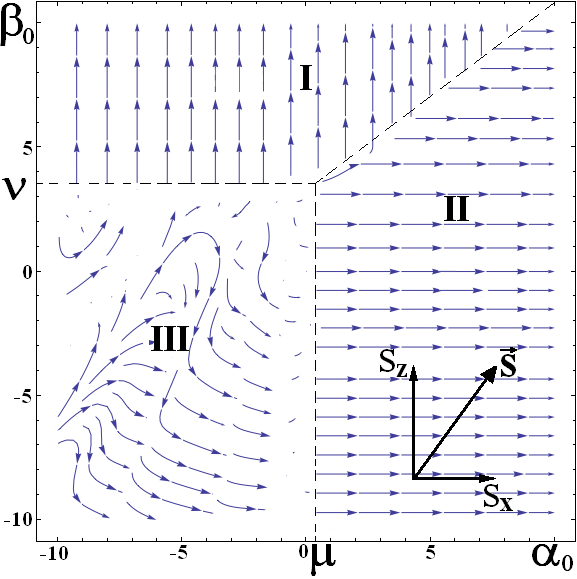}

\caption{Molecular dipoles orientation $(s_{x}(0),s_{z}(0))$ at the interface
boundary depending on the surface interaction parameters $(\alpha_{0},\beta_{0})$.\label{fig:Three types of surfaces }}
\end{figure}

To understand the properties of the liquid interfaces, consider the
semi-infinite water sample resided in the region $z>0$, contacting
an infinite plane surface $z=0$. Since there is no external electric
field in the system, $\mathbf{E}=\mathbf{E}_{P}=(0,0,-4\pi P_{0}s_{z})$.
The mean field solution is obtained via the free energy minimum, using
the trial function in the form $\boldsymbol{s}=(s_{x}^{(1)},0,s_{z}^{(1)})e^{-z/L_{T}}+(s_{x}^{(2)},0,s_{z}^{(2)})e^{-z/R_{D}}$,
where $s_{x,y}^{(1,2)}$ are the four variable parameters. The results
of the minimization are represented on Figure \ref{fig:Three types of surfaces }. 

We find three distinctly different types of the solutions, depending
on the properties of the surface (the parameters $\alpha_{0}$ and
$\beta_{0}$) . If $\beta_{0}$ is sufficiently large, in region $I$,
then the water molecules are polarized along the normal to the interface
surface, $|s_{x}^{(1,2)}|\approx0$, which corresponds to hydrophilic
property. Moreover, $|s_{z}^{(1)}|\approx0$ and the polarization
of the liquid extends exponentially into the liquid $|s_{z}|\sim e^{-z/R_{D}}$.
In region $II$, the water molecules are polarized along the interface
surface, $|s_{z}^{(1,2)}|,|s_{x}^{(2)}|\approx0$ and $|s_{x}^{(1)}|\sim e^{-z/L_{T}}$,
which is exactly what we expect from a hydrophobic surface \cite{kohlmeyer1998orientational}.
In region $III$, the variational solution vanishes and the polarization
of the liquid can exist only due to the thermal fluctuations. 

Boundaries between the three regions can be found in an analytical
form, due to the exceptional simplicity of the mean field solutions,
using the trial function $\boldsymbol{s}=(s_{x}e^{-z/L_{T}},0,s_{z}e^{-z/R_{D}})$,
so that $G=SP_{0}^{2}\sqrt{C}\cdot(R_{2}+BR_{4})$, 
\[
R_{2}=\frac{1}{2}\left(\mu-\alpha_{0}\right)s_{x}^{2}+\frac{1}{2}\left(\nu-\beta_{0}\right)s_{z}^{2}+\frac{A}{\mu+\nu}s_{x}s_{z},
\]
\[
R_{4}=\frac{s_{x}^{4}}{4\mu}+\frac{4s_{x}^{3}s_{z}}{3\mu+\nu}+\frac{3s_{x}^{2}s_{z}^{2}}{\mu+\nu}+\frac{4s_{x}s_{z}^{3}}{\mu+3\nu}+\frac{s_{z}^{4}}{4\nu}>0,
\]
where $\mu=\sqrt{A}$, and $\nu=\sqrt{4\pi+A}$. The minimization
of $G$ with respect to $s_{x,y}$ shows that the interface is hydrophobic
(Region $I$) if $\beta_{0}>\nu$ ($\alpha_{0}<\mu$) or $\beta_{0}>\beta(\alpha_{0})=\nu+\left(\alpha_{0}-\mu\right)\sqrt{\mu/\nu}$
($\alpha_{0}>\mu$). The wetting energy is $G_{I}=-SP_{0}^{2}\sqrt{C}\nu\left(\beta_{0}-\nu\right)^{2}/4B$.
The hydrophobic type of the interface (Region $II$) corresponds to
$\alpha_{0}>\mu$ and $\beta_{0}<\beta(\alpha_{0})$, where $G_{II}=-SP_{0}^{2}\sqrt{C}\mu\left(\alpha_{0}-\mu\right)^{2}/4B$.
The fluctuation dominated Region $III$ corresponds to $\alpha_{0}<\mu$
and $\beta_{0}<\nu$, when the mean field $G_{III}=0$.

Interaction forces between plane surfaces in water for Regions $I$
and $II$ can be calculated using the same formalism. Consider first
two hydrophilic bodies (e.g. biological membranes) with plane surfaces
at $z=\pm d/2,$ separated by the water filled layer of width $d$.
In extreme hydrophilic case, $\alpha_{0}\approx0$ and $\beta_{0}\gg1$,
the mean field solution gives $s_{z}(\pm d/2)=\pm s_{0}$, where $s_{0}\alt1$
and the minimization of $G$ recovers the experimentally observed
dependence (\ref{eq: The disjoining pressure}) with $P_{1}=2\pi P_{0}^{2}s_{0}^{2}\approx3\cdot10^{10}\, dyn/cm^{2}$
and $\lambda=R_{D}$. Similarly in Region $II$, for $d\agt R_{D}$
we obtain $|s_{z}|\approx0$ and $P=-P_{2}e^{-d/\lambda}$, where
$P_{2}\sim P_{1}\sim2\pi P_{0}^{2}$ and $\lambda\sim L_{T}$ in agreement
with the experimental data \cite{pashley1985attractive,claesson1988very}.
Therefore, in both cases the interaction force decays exponentially
with the distance between the planes. Both the decay length and the
pre-exponential factor depend on the properties of the surface material. 

\begin{figure}
\includegraphics[width=0.9\columnwidth]{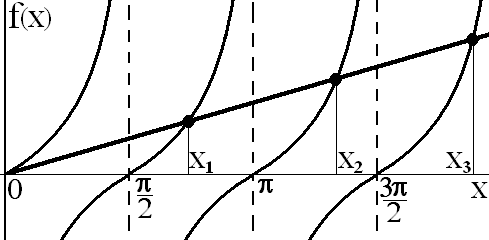}

\caption{Graphical solution of the characteristic equation for $q_{m}$. \label{fig: Graphical solution of the quantization equation} }
\end{figure}

Region III represents a very special case, where the mean field polarization
vanishes and the energy of the liquid is determined by thermal fluctuations.
The geometry dependent part of the free energy leads to the interaction
between the boundaries exactly in the way electromagnetic field fluctuations
lead to appearance of the Casimir forces \cite{casimir1948attraction}.
To describe the fluctuations, we use Eqs. (\ref{eq:GB})-(\ref{eq:GS}),
keeping only terms $\sim s^{2}$. The free energy of the liquid takes
a form $G=\left\langle \boldsymbol{s}\left|\widehat{H}\right|\boldsymbol{s}\right\rangle $,
where $\widehat{H}$ is a properly constructed self-conjugated operator.
Diagonalization is produced by the decomposition $\boldsymbol{s}=\sum_{n}c_{n}\boldsymbol{s}_{n}$
over the complete set of orthogonal and normalized eigen-mode functions
$\{\boldsymbol{s}_{n}\}$, so that $G=\sum_{n}k_{n}^{2}|c_{n}|^{2}$,
where $k_{n}^{2}$ are the eigen-numbers of $\widehat{H}$ corresponding
to the modes $\boldsymbol{s}_{n}$ and enumerated by the index $n$.
Each $c_{n}$ is an independent variable.

In the case of the two plane surfaces separated by the distance $d$,
the liquid is translational invariant along the interfaces surfaces
and the mode functions can be represented in the form $\boldsymbol{s}_{n}=\left(u_{n}\left(z\right),0,v_{n}\left(z\right)\right)\exp e^{ipx}$.
The solutions are characterized by the set of numbers $n=\left(\boldsymbol{p},P,m\right)$,
where $\boldsymbol{p}$ is the two-dimensional wave vector and $P=\pm1$
is the parity of the function, $u_{n}\left(-z\right)=Pu_{n}\left(z\right)$.
Depending on the parity, the mode functions $u_{n}(z)$ and $v_{n}(z)$
are the linear combinations of $\sin q_{m}^{(1,2)}z$ and $\cos q_{m}^{(1,2)}z$,
where $q_{m}^{(1)}=\sqrt{k_{n}^{2}-p^{2}-A}$ and $q_{m}^{(2)}=\sqrt{q_{m}^{2}-4\pi}$.
In the practically important case $d\agt R_{D}$, all the terms containing
$q_{m}^{(2)}$ are small, $\sim\exp(-d/R_{D})$, and as such can not
contribute to the interaction force. This is because at sufficiently
large distances from the surfaces only the so-called ``force-less'',
$\boldsymbol{E}_{P}=0$, fluctuations of the liquid contribute to
thermodynamic functions \cite{fedichev2006long}. The wave vector
$q_{m}\equiv q_{m}^{(1)}$ is the solution of the characteristic equation:
$f\left(X_{m}\right)=aX_{m}$, where $X_{m}=q_{m}d/(2\sqrt{C})>0$,
and $a=2\sqrt{C}/(\alpha_{0}d)$. The function $f\left(X\right)$
has the period $\pi/2$ and is obtained by a periodic shift of $\pi/2$
of the main branch of tangent function, $\tan X$ for $-\pi/2<X<\pi/2$,
as shown on Figure \ref{fig: Graphical solution of the quantization equation}. 

The equilibrium free energy of the system is 

\[
G_{eq}\left(d\right)=-T\ln\int D\mathbf{s}(\mathbf{r})e^{-G/T}=\frac{TS}{C}\int\frac{d^{2}p}{\left(2\pi\right)^{2}}\sum_{m=1}^{\infty}\ln k_{m},
\]
and, exactly as in the calculation of Casimir energy, formally diverges.
To compute the sum, we formally write
\[
G_{eq}\left(d\right)=-\frac{TS}{2\pi C}\begin{array}[t]{c}
\lim\\
\varepsilon\rightarrow0
\end{array}\intop_{0}^{\infty}pdp\sum_{m=1}^{\infty}\intop_{1}^{\infty}\frac{dx}{x}e^{-\varepsilon k_{m}x}+{\rm Const}
\]
Now we can follow \cite{boyer1970quantum} and perform the summation
using the Cauchy's argument principle. After the regularization, $G_{eq}\left(d\right)\rightarrow G_{eq}\left(d\right)-G_{eq}\left(\infty\right),$
we derive the expression:
\[
G_{eq}\left(d\right)=\frac{TS}{4\pi L_{T}^{2}}\intop_{1}^{\infty}tdt\ln\left[\frac{\left(\lambda t-\tanh y_{0}t\right)\left|\tanh y_{0}t-\frac{1}{\lambda t}\right|}{\left(\lambda t-1\right)\left|1-\frac{1}{\lambda t}\right|}\right],
\]
where $y_{0}=d/2L_{T}$ is the dimensionless distance, and $\lambda$
is a material dependent quantity (in Region III $\mu/\alpha_{0}>1$).
In the two most important limiting cases, the interaction energy takes
form 
\[
G_{eq}\left(d\right)\approx-\frac{TS}{2\pi L_{T}d}\exp\left(-\frac{d}{L_{T}}\right),\; d\gg L_{T},
\]
\[
G_{eq}\left(d\right)\approx-0.53\frac{TS}{\pi d^{2}},\; d\ll L_{T},
\]
corresponding to the attraction. This means that the interaction is
universal, the dependence on the material constants is weak and can
only be found at intermediate distances $d\sim L_{T}$. 

The attraction of hydrophobic bodies in our model has an entropic
nature, in accordance with earlier predictions \cite{huang2001scaling,chandler2005interfaces}.
Hydrophobic surfaces order molecules of water and the effects of the
molecular ordering in (\ref{eq:GB}) manifests itself in two different
ways: from the entropy contribution to $V(\mathbf{s}^{2})$, as in
our case, or through the Oseen energy term, modeling the hydrogen
bonding. The latter describes the short-range forces and decays at
distances $\sim R_{D}$. Hence, for small bodies of sizes $\alt R_{D}$
the hydration energy is proportional to the volume \cite{chandler2005interfaces,huang2001scaling}.
The longer range interaction between hydrophobic bodies at distances
$\sim L_{T}$ originates from the long-range dipole-dipole interaction
between the molecules and thus requires a complete model like (\ref{eq:GB})-(\ref{eq:GS}),
which naturally includes both distance scales.

On a side note, we predict that under specific conditions there could
be a special limit. When the liquid interfaces fail to polarize water
molecules, the fluctuations of molecular polarization become strong
and the interaction becomes very weak but attractive. The fluctuations
may also be relevant next to hydrophobic interfaces, where the mean
field ordered state of the liquid may break in a BKT-like phase transition
\cite{men2009nature,vasiliev2013universality}, which have been observed
in molecular dynamics calculations for the hydration water layers
\cite{oleinikova2005percolation}. 

In summary, we find the polar liquid phenomenology (\ref{eq:GB})-(\ref{eq:GS})
earlier proposed \cite{fedichev2006long,men2011possible,men2009nature}
paints a very physically rich picture of possible wetting regimes
and interactions. If used to calculate the interactions between the
hydrophilic planes, the model can be considered as the natural improvement
of ideas \cite{marcelja1976repulsion,belaya1986hydration,Ramirez,gong2009langevin,azuara2008incorporating,koehl2009beyond,beglov1997integral}.
Our model generalizes the order parameter (the net polarization of
the water molecules) in a form useful both for hydrophobic and hydrophilic
interfaces, correctly predicts the sign and the distance dependence
of the interaction forces depending on the properties of the surfaces.
We show that the interactions originate from the fundamentally non-linear
and non-local polarizability of the liquid. Our model is characterized
by two scales, $R_{D}$ and $L_{T}$, instead of a single scale $\sim R_{D}$
from \cite{belaya1986hydration,kornyshev1989fluctuation,leikin1990theory}.
The experimental observation of both decay lengths in \cite{leneveu1976measurement}
and \cite{pashley1985attractive,claesson1988very} together with the
prediction \cite{men2011possible,fedichev2013application} and subsequent
experimental observation of the ferroelectric features in the bulk
liquid water near the $\lambda-$point \cite{fedichev2011experimental,bordonskiy2012study,bordonskiy2013electric}
shows both general consistency and applicability of the model for
realistic calculations of macroscopic bodies interactions in water.
This observation makes our model the minimal continuous model capable
of predicting finer effects depending both on the hydrogen-bond network
properties and the electrostatic interactions of the water molecules. 

The authors are grateful to Prof. V.G. Levadny for the fruitful discussions.
Quantum Pharmaceuticals supported this work.

\bibliographystyle{apsrev}
\bibliography{../Qrefs}

\end{document}